\documentstyle[12pt]{article}
\newcommand{\al}{\alpha}             \newcommand{\bt}{\beta}
\newcommand{\dl}{\delta}			\newcommand{\la}{\Lambda}

\newcommand{\ep}{\epsilon}           \newcommand{\G}{\Gamma}
\newcommand{\pa}{\partial}			\newcommand{\om}{\Omega}
\newcommand{\ty}{\tilde{y}}   \newcommand{\tz}{\tilde{z}}
\newcommand{\tk}{\tilde{k}} 	
 	
\newcommand{\be}{\begin{equation}}	\newcommand{\cL}{\cal L}
\newcommand{\cG}{\cal G}		\newcommand{\cB}{\cal B}
\newcommand{\ee}{\end{equation}}
\newcommand{\bq}{\begin{eqnarray}}
\newcommand{\eq}{\end{eqnarray}}
\begin{document}
\baselineskip= 24 truept
\begin{titlepage}
\title { Gauging of T-Duality Invariant Worldsheet String Actions}
\author{\sc Alok Kumar  \\
\\
 Institute of Physics, Bhubaneswar-751 005, India. \\
 email: kumar@iopb.ernet.in }
\date{}
\maketitle
\thispagestyle{empty}
\vskip .6in
\begin{abstract}
\vskip .2in
T-Duality invariant worldsheet string actions, recently written down by
Schwarz and Sen, are coupled to the worldsheet gauge fields.
It  is shown that the
integration of the dual coordinates gives the conventional,
vector, axial and chiral, gauged string actions for the appropriate
choice of the gauged isometries. Alternatively, the gauge field
integration is shown to give a T-duality invariant action
which matches with the corresponding results  known earlier.

\end{abstract}
\bigskip
\flushright {IP/BBSR/93-32}
\vfil
\end{titlepage}
\eject

It is known that the string effective action is invariant under an
$O(d, d)$ group \cite{duff}\cite{ven}\cite{sen}
\cite{giveon} of symmetry transformations.
Its discrete subgroup, known as  {\it T-Duality}\cite{sch2},
interchanges Kaluza-Klein and winding-mode excitations.
Such symmetries are expected to play an important role in
classifying the string vacua \cite{maha}.
Therefore  it is important to explore whether T-duality is
a symmetry of the full string theory, rather than the string
effective action only. As a step in this direction,
Schwarz and Sen \cite{sch2}
have recently written down a worldsheet string action with explicit
T-duality symmetry by introducing a set of extra (dual)
coordinates on the worldsheet. By integrating out the dual
coordinates, one obtains the conventional  string action.

In a parallel development, recently, there has been a great deal of
interest in obtaining classical solutions of string theory
from the gauged string actions \cite{witten}
such as the gauged WZW models.
This is achieved by gauging the
abelian \cite{giveon}\cite{witten}\cite{rocek} or nonabelian \cite{quevedo}
"chiral"-isometries of the string backgrounds through the coupling of
nondynamical
gauge fields to the corresponding conserved currents.
The gauge field integration gives a string action which
represents an interesting class of classical  string backgrounds such
as 2D blackholes \cite{witten}, 3D charged black strings
\cite{hor} etc.

In this paper, it is shown that the
worldsheet gauge fields can be
coupled directly to the T-Duality invariant Schwarz-Sen action. Using
the conditions for the existence of chiral conserved currents
for this action, which turn out to be
identical to the ones for the usual string actions \cite{rocek},
the invariance of the gauged action is demonstrated.
The integration of the dual coordinates now gives
directly the conventional {\it gauged}
string actions \cite{giveon}\cite{rocek}\cite{kar} for the vector, axial or
chiral \cite{tye} gauging depending on the  choice of
the gauged subgroup.
Our action has a similar form as the Schwarz-Sen action for
the heterotic string \cite{sch2}. However the
field content as well as the symmetries in the two cases are quite
different.

We also show, through an example,
that when the gauge fields are integrated out, instead of
the dual coordinates, one obtains a worldsheet action which is also of
Schwarz-Sen form and the resulting background matches with the ones
obtained before.

We begin by writing down the T-duality invariant
action for the background
metric and antisymmetric tensor of the form \cite{ven}
\be
G\; =\;\left (\matrix {{-1} & {0} \cr
	{0} & {{\cG}_{ij}} \cr }\right ),
	\;\;\;\;\;\;
B\; =\;\left (\matrix {{0} & {0} \cr
			{0} & {{\cB}_{ij}} \cr }\right ).	\label{back}
\ee
This background
is dependent only on a single coordinate
$x$ and is independent of the rest of the $d$
coordinates $y^i$'s. In
this paper we are restricting $G$ and $B$ to the above
form. Although more general cases can also be handled on similar lines.
In the language of ref.\cite{sch2}, the above form for $G$ and $B$
implies
that we have ignored the "space-time" gauge fields.
By introducing $d$ extra coordinates $\ty^i$'s, called the {\it dual
coordinates}, and by defining
$	y^a \equiv \left( \matrix{ {y^i} \cr {\ty^i}} \right),$
the ungauged T-duality invariant Lagrangian is written as:
\be
	{\cL} = -{1\over 2} \eta^{\al\bt}\pa_{\al}x\pa_{\bt}x
	- {1\over 2} \pa_0 y^a L_{a b} \pa_1 y^b
	- {1\over 2} \pa_1 y^a {({\it LML})}_{ab} \pa_1 y^b ,  \label{tdual}
\ee
where
\be
M\; =\;\left (\matrix {{{\cG}^{-1}} & {-{\cG}^{-1}{\cB}} \cr
		{{\cB} {\cG}^{-1}} & {{\cG}- {\cB}{\cG}^{-1}{\cB}} \cr }\right ),
						\;\;\;\;\;\;
				L\; = \;\left(\matrix{ {0} & {I_d} \cr
						{I_d} & {0}		\cr}	\right),
\ee
$I_d$ is the $d$-dimensional identity matrix
and $(0, 1)$ are respectively the $(\tau, \sigma)$ worldsheet coordinates.
This action has two dimensional Lorentz invariance, although it is not
explicit\cite{sch2}\cite{sch1}. By addition of terms which vanish in  the
orthonormal gauge, it can also be made local reparameterization and Weyl
invariant. The T-duality symmetry is manifest and the corresponding
transformations are,
$y^a  \rightarrow {{\om}^a}_b y^b $ and $M  \rightarrow {\om} M {\om}^T $.
The conventional (ungauged) string action
is now obtained by integrating out $\ty$.
Classically this can be done by using the equation of motion
for $y^i$'s derived from (\ref{tdual}):
\be
\pa_1 \left[ \pa_0 y - {\cG}^{-1}{\cB} \pa_1 y +
				{\cG}^{-1} \pa_1 \tilde{y}
				\right] = 0,	\label{emo}
\ee
where $y\equiv \left( y^i \right)$ $\ty\equiv \left( \ty^i \right)$
are $d\times 1$ column vectors.
By using the freedom of $\tau$-dependent translation of $y$ in
eqn.(\ref{emo}), it can be simplified to a first order
constraint equation for $\ty$:
\be
	\pa_1 \tilde{y} = - {\cG} \pa_0 y + {\cB} \pa_1 y.	\label{constraint}
\ee
Then by substituting for $\ty$ from (\ref{constraint}) into
the Lagrangian (\ref{tdual}) we obtain:
\be
{\cL} = -{1\over 2} \eta^{\al \bt} \pa_{\al}x\pa_{\bt}x
	+ {1\over 2} {\cG}_{ij}\eta^{\al \bt} \pa_{\al}y^i \pa_{\bt}y^j
	- {1\over 2} {\cB}_{ij}\epsilon^{\al \bt} \pa_{\al}y^i \pa_{\bt}y^j
\ee
which is just the conventional ungauged string
Lagrangian in the background eqn.(\ref{back}).

Now, to couple the worldsheet gauge fields to the T-duality
invariant action,
we first analyze its symmetries. The Lagrangian in (\ref{tdual}) is
invariant under ($\sigma$, $\tau$)-independent translations in $y^a$'s.
The conserved currents corresponding to a (constant) Killing vector,
$K^a\; \equiv \;\left (\matrix{ {k^i} \cr {\tilde{k^i}}
		\cr }\right )$, are
\bq
	J_0 &= -{1\over 2}K^a L_{ab} \pa_1 y^b	\cr
	J_1 &= {1\over 2} K^a {(LML)}_{ab} \pa_1 y^b, \cr		\label{current}
\eq
and satisfy the conservation law,
$\pa_0 J_0 - \pa_1 J_1 = 0 $. By defining $J_{\pm} = J_0 \pm J_1$ and
$\pa_{\pm} = \pa_0 \pm \pa_1$, the conservation law
can be written as $\pa_- J_+ + \pa_+ J_- = 0$, and from (\ref{current}) we
have
\be
	J_+  = -{1\over 2}\left[ k^T (-{\cG}+ {\cB} {\cG}^{-1}{\cB})
		+ {\tk}^T(I+{\cG}^{-1}{\cB}) \right]\pa_1 y
		- {1\over 2} \left[ k^T (I - {\cB} {\cG}^{-1} )
				- {\tk}^T {\cG}^{-1} \right]\pa_1 \ty. \label{jplus}
\ee
\be
	J_-  = -{1\over 2}\left[ k^T ({\cG}- {\cB} {\cG}^{-1}{\cB})
		+ {\tk}^T(I-{\cG}^{-1}{\cB}) \right]\pa_1 y
		- {1\over 2} \left[ k^T (I + {\cB} {\cG}^{-1})
				+ {\tk}^T {\cG}^{-1} \right]\pa_1 \ty .
				\label{jminus}
\ee
The following relations are therefore derived for  the
existence of chiral conserved currents in the T-duality invariant
string action:
\bq
	\tk_L &= ({\cG}+{\cB})k_L, \;\;\;\;\;\;\;\;for\;\;J_+ = 0 \cr
	\tk_R &= -({\cG}-{\cB})k_R, \;\;\;\;\;\;\;\;for\;\;\;J_- = 0 \cr
									\label{iso}
\eq
However eqns.(\ref{iso}) are consistent with our original assumption
of isometries being $\sigma, \tau$-independent
translations only if the conditions:
\bq
	{(\dot{\cG}+\dot{\cB})k_L} = 0, \cr
	{(\dot{\cG}-\dot{\cB})k_R} = 0, \cr			\label{cond}
\eq
are satisfied, where dots denote the derivative with respect to $x$.
Interestingly, these conditions are
precisely the ones written down in ref.\cite{rocek} for the existence of chiral
isometries in the conventional string action, i.e.
\be
	\pa_{\mu} K_{\nu} - \G^{\rho}_{\mu\nu} \pm
		{1\over 2} H^{\rho}_{\mu\nu} = 0,
\ee
when one restricts to the background of the form in eqns.(\ref{back}).
The expressions for the
currents also match with the ones in \cite{rocek} after
the substitution for $\ty$ from eqn.(\ref{constraint}). We have from
(\ref{jplus}) and (\ref{jminus}),
\bq
	J^L_- &= k_L^T {\cG} \pa_-y,\;\;\;\;\;\;\;J^L_+ = 0, 	\cr
	J^R_+ &= k_R^T {\cG} \pa_+y,\;\;\;\;\;\;\;J^R_- = 0.	\cr
								\label{chiral}
\eq

Therefore, given solutions $k_L$ and $k_R$ of eqns.(\ref{cond}),
the left and right-chiral isometries for the T-duality invariant
action are written as:
\be
K_L\; = \;\left (\matrix {{k_L} \cr {} \cr
			{({\cG}+{\cB})k_L} \cr }\right ),
		\;\;\;\;\;\;
K_R\; \equiv \;\left (\matrix{ {k_R} \cr {} \cr {-({\cG}-{\cB})k_R}
		\cr }\right ).			\label{bklkr}
\ee
In general, in a given theory, there may exist several such solutions.
However, we deal with only one of the left and right isometries.

We first consider the vector gauging of the T-duality invariant
action. The gauged isometry is then written as,
\be
	K_V = K_L + K_R = \left(\matrix{ {k_V} \cr {{\cG} k_A +
						{\cB} k_V } }\right),
									\label{kv}
\ee
where $k_V = k_L+k_R$ and $k_A = k_L - k_R$. By defining,
$D_{\al}y^a = \pa_{\al}y^a + A_{\al}K_V^a$,
the gauge invariant action is given by,
\bq
	{\cL} = -{1\over 2} \eta^{\al\bt}\pa_{\al}x\pa_{\bt}x
	- {1\over 2} D_0 y^a L_{a b} D_1 y^b
	- {1\over 2} D_1 y^a {({\it LML})}_{ab}D_1 y^b  \cr
	- {1\over 2}\epsilon^{\al\bt}
	A_{\al}K_V^a L_{ab}\pa_{\bt}y^b \cr			\label{gaction}
\eq
The invariance
of the first three terms in the action under the local transformations,
	$\dl y^a = \la (\sigma, \tau) K_V^a$ and
	$\dl A_{\al} = - \pa_{\al}\la (\sigma, \tau)$
is rather obvious. To prove that the last term
in (\ref{gaction}) is also gauge
invariant, we note that since $k_L$ and $k_R$ satisfy eqns.(\ref{cond}),
$k_L^T {\cG} k_L$ and $k_R^T {\cG} k_R$ are
constants and can be set equal by normalizing $k_L$ and $k_R$
appropriately. Then since, upto total derivative,
the  variation of the last term in
eqn.(\ref{gaction}) is given by
\be
	-{1\over 2}\dl \left(\epsilon^{\al \bt}
		A_{\al}K_V^a L_{ab}\pa_{\bt}y^b \right)
		= -{1\over 4}\ep^{\al \bt} F_{\al\bt}\la (\sigma, \tau)
		(K_V^a L_{ab}K_V^b),
\ee
which vanishes by using
\be
	(K_V^a L_{ab}K_V^b) = 2{(k_L^T {\cG} k_L
						- k_R^T {\cG} k_R)} = 0,	\label{norm}
\ee
hence the full action (\ref{gaction}) is gauge
invariant. To summarize, the
construction of gauge and T-duality invariant string action
requires obtaining a solution for $k_L$ and $k_R$ from
eqn.(\ref{cond}) which
are normalized so that eqn.(\ref{norm}) is satisfied. The gauge invariant
action for the vector case is then given by eqn.(\ref{gaction}).

The conventional vector gauged string action
\cite{giveon}\cite{rocek} is now obtained from
the T-duality invariant gauged action, eqn.(\ref{gaction}), by
using the constraint equation,
\be
	D_1 \tilde{y} = - {\cG} D_0 y + {\cB} D_1 y,	\label{motion}
\ee
which is obtained in the same way as (\ref{constraint}) and
can be used to eliminate $\ty$ in favour of $A_{\al}$
and $y$. As a result we find,
\bq
	{\cL} = -{1\over 2} \eta^{\al\bt}\pa_{\al}x\pa_{\bt}x
		+ {1\over 2} \pa_0 y^T {\cG} \pa_0 y -
		{1\over 2} \pa_1y^T {\cG} \pa_1 y
		- \pa_0 y^T {\cB} \pa_1 y     	\cr
		- A_1 k_V^T {\cG} \pa_1y  + A_0 k_V^T {\cG} \pa_0y
		- A_0 k_V^T {\cB} \pa_1 y - \pa_0 y^T {\cB} k_V A_1 	\cr
			-A_0 \tk^T_V \pa_1 y + A_1\tk_V^T \pa_0 y
		+ {1\over 2} A_0^2k_V^T {\cG} k_V
		- {1\over 2} A_1^2k_V^T {\cG} k_V		\cr
		- A_0 A_1 k_V^T {\cB} k_V
		+ {1\over 2} A_0 A_1 k_V^T \tk_V
		- {1\over 2} A_0 A_1 \tk_V^T k_V. \cr
											\label{eliminate}
\eq
The last three terms in this action cancel out. But they have
been kept for the discussion later on of the chiral gauging case.
Then using explicit forms of $\tk_V$ from eqn.(\ref{kv})
one finds that the gauge fields, $A_{\pm} = A_0 \pm A_1$, couple directly to
the chiral conserved currents (\ref{chiral}) and the
Lagrangian (\ref{eliminate}) simplifies to,
\bq
	{\cL} = -{1\over 2}\eta^{\al\bt}\pa_{\al}x\pa_{\bt}x
	+ {1\over 2} \pa_+y^T({\cG}+{\cB})\pa_-y
	+ A_+ (k_L^T {\cG} \pa_-y)  \cr
	+ A_-(k_R^T{\cG}\pa_+y) + {1\over 2} A_+A_- (k_V^T {\cG} k_V).	\cr
									\label{simple}
\eq

Now, to compare it with the known results of the vector gauged actions,
we mention the models discussed by Rocek and Verlinde \cite{rocek}.
By setting the cross terms in $x$ and $y$, i.e.
$G^{L, R}_a$, in that reference to zero,
the metric and antisymmetric tensor are specified by the matrices,
\be
	\cG = \left(\matrix{ 1 & b \cr
				   b &1 \cr		}\right)
		\;\;\;\;\;
	\cB = \left(\matrix{ 0& -b \cr
				   b & 0 \cr		}\right),	\label{gb}
\ee
and the chiral isometries for this background are,
\be
	k_L = \left( \matrix{ 0 \cr 1 \cr }\right)
		\;\;\;\;\;\;
	k_R = \left( \matrix{ 1 \cr 0 \cr }\right).		\label{klkr}
\ee
When these explicit form of $\cG$, $\cB$, $k_L$ and $k_R$ is
substituted into (\ref{simple}) we get, after a field
redefinition, $A_{\pm} \rightarrow {1\over
2}A_{\pm}$,
\bq
	{\cL} = {1\over 2}[ -\eta^{\al\bt} \pa_{\al}x \pa_{\bt}x
			+ \pa_+y_1 \pa_-y_1 + \pa_+y_2 \pa_-y_2
			+ 2b\; \pa_+y_2 \pa_-y_1			\cr
			+ A_+ (\pa_-y_2 + b \pa_-y_1)
			+ A_- (\pa_+y_1 + b \pa_+y_2)
			+{1\over 2} A_+A_- (1+b)]		\cr
\eq
which matches with eqn.(3.10) of \cite{rocek} for $G^{L, R}_a = 0$.

We have thus shown
that the vector gauging of the string action can be done direclty for the
T-duality invariant action and the integration of the dual
coordinates reproduces
the conventional vector gauged string actions. Repeating
this exercise for the axial
gauging, i.e. for the isometry $K_A = K_L - K_R$, one obtains an
expression similar to eqn.(\ref{simple}) with the replacement
$k_R \rightarrow - k_R$ and $k_V \rightarrow k_A$.
Once again the result matches with ref.\cite{rocek}.

Recently, in an interesting paper,
a classification of possible gaugings in two dimensions
was done by Chung and Tye \cite{tye}. It was shown that, apart
from the vector and axial gaugings, there is another
consistent gauging in two dimensions, namely the {\it chiral gauging}.
This is characterized by independent left and
right gauge invariance and requires the presence of two gauge fields
$A^L$ and $A^R$. The anomaly free gauge choice is given by,
$A^L_- = A^R_+ = 0$.

The T-duality invariant {\it chiral} gauged string action
has the same form as (\ref{gaction}) with the
replacement, $D_{\al} y^a \rightarrow D^c_{\al} y^a$ and
$A_{\al}K_V^a \rightarrow A^L_{\al}K_L^a + A^R_{\al}K_R^a$, where,
\be
D^c_{\al}y^a = \pa_{\al}y^a + A^L_{\al}K_L^a + A^R_{\al}K_R^a.
\ee

The symmetry transformations are now written as,
	$\dl y^a = \la^L (\sigma^+) K_L^a + \la^R (\sigma^-) K_R^a$,
$\;$	$\dl A^L_+ = - \pa_+\la^L (\sigma^+)$ and
	$\dl A^R_- = - \pa_-\la^R (\sigma^-)$, where
	$\sigma^{\pm} = \tau \pm \sigma$.
Then using $\dl D_+ y^a = \dl D_- y^a = 0$,
the gauge invariance of the chiral gauged
action can be shown if one also uses
$K_L^a L_{a b} K_R^b = 0$ for $K_L$ and $K_R$ in (\ref{bklkr}).

The coordinates $\ty$ can once again be eliminated by using the
equations of motion which are of the form (\ref{motion}), with
$D_{\al} \rightarrow D^c_{\al}$. After the substitution for $\ty$ in
terms of $A^L_{\al}$, $A^R_{\al}$ and $y$, the action has the
form (\ref{eliminate}) with substitutions
$A_{\al} k_V \rightarrow A^L_{\al}k_L + A^R_{\al}k_R$
and $A_{\al} \tk_V \rightarrow
A^L_{\al}\tk_L + A^R_{\al}\tk_R$.
Finally by writing $\tk$ in terms of $k$ we get, for the  chiral gauging,
\bq
	{\cL} = -{1\over 2} \eta^{\al\bt}\pa_{\al}x\pa_{\bt}x
	+ {1\over 2} \pa_+y^T({\cG}+{\cB})\pa_-y
	+ A^L_+ (k_L^T {\cG} \pa_-y) \cr
	+ A^R_-(k_R^T{\cG}\pa_+y)
	+  A^L_+A^R_- (k_L^T {\cG} k_R).	\cr \label{cgauge}
\eq
Once again, to compare with the known results we note that, for the
backgrounds of the type considered in eqn.(\ref{gb}),
the Lagrangian (\ref{cgauge}) can be rewritten, by a field
redefinition $A_{\pm} \rightarrow {1\over 2} A_{\pm}$, as
\bq
	{\cL} = {1\over 2}\left[-\eta^{\al\bt}\pa_{\al}x\pa_{\bt}x
			+ \pa_+y_1 \pa_-y_1 + \pa_+y_2 \pa_-y_2
			+ 2b\; \pa_+y_2 \pa_-y_1		\right.\cr
			+ A^L_+ (\pa_- y_2 + b \pa_-y_1 )
			+ A^R_- (\pa_+y_1 + b \pa_+y_2)
\left.			+{1\over 2} A^L_+A^R_- \;b \right]
\eq
which matches with eqn.(7) of \cite{kar} when one sets $b = coshx$.

After presenting the gauge and T-duality invariant string actions and
showing their connection to the conventional gauged actions, we now
show, through an example of vector gauging,
that when gauge fields are integrated
out, keeping both $y$ and $\ty$, we get an action which is again of
the form (\ref{tdual}). For this example, we show the
decoupling of a pair of coordinates. As a result, the final action
is a lower dimensional one and gives the 2D black hole
background when the ungauged action is written for the background of
the $SL(2, R)$ WZW model.

For the gauge field
integration we write down the vector gauged
Lagrangian (\ref{gaction}) explicitly as,
\bq
	{\cL} = -{1\over 2}\eta^{\al \bt}\pa_{\al}x \pa_{\bt} x
		-{1\over 2}\pa_0 y^a L_{ab}\pa_1 y^b
		-{1\over 2}\pa_1y^a {(LML)}_{ab}\pa_1 y^b	\cr
		- A_0 (K_V^a L_{ab}\pa_1 y^b)
		- A_1 \left(K_V^a {(LML)}_{ab}\pa_1 y^b\right)
		-{1\over 2} A_1^2 [ K_V^a {(LML)}_{ab}K_V^b].	\cr
										\label{explicit}
\eq
The gauge symmetry of the action allowes us  to make
a gauge choice, $K_V^a L_{ab} y^b = 0, $ which is similar to the one
used in the gauging of conventional string actions \cite{witten}\cite{hor}.
One of the coordinates is eliminated by this gauge choice.
However T-duality symmetry then requires the decoupling of one more
coordinate. We now show that indeed there is one more combination
which decouples.

For our gauge choice, the Lagrangian (\ref{explicit}) can be written
after integrating out $A_1$ by its equation of motion as,
\bqu
	{\cL} = -{1\over 2}\eta^{\al \bt}\pa_{\al}x \pa_{\bt}x
		-{1\over 2}\pa_0 y^a L_{ab}\pa_1 y^b
		-{1\over 2}\pa_1y^a {(LML)}_{ab}\pa_1 y^b	\cr
		+  {1\over 2K_V^2} {[ K_V^a {(LML)}_{ab}\pa_1 y^b]}^2	\cr
										\label{lint}
\eq
where $K_V^2 = K_V^T (LML) K_V$. We now restrict to the specific choice
of the background (\ref{gb}).
For this example, it can be shown that there is a set of four basis
vectors, $K_i \equiv$ $(K_V$, $K_A$, $K_M$, and $K_N)$
which satisfy the orthogonality and completeness relations:
\be
	{K_i}^T {(LML)} K_j = 0 \;\;\;\;\;\;\;\; i \neq j	\label{ortho}
\ee
\be
{(LML)}\left[{K_V K_V^T\over K_V^2} + {K_A K_A^T\over K_A^2}
		+ {K_M K_M^T\over K_M^2}
		+ {K_N K_N^T\over K_N^2} \right] = I,		\label{comp}
\ee
where $K_i^2 = K_i^T LML K_i$, and $I$ is a $4$-dimensional identity
matrix. Using (\ref{comp}), the Lagrangian (\ref{lint}) can be written as,
\be
	{\cL} = -{1\over 2}\eta^{\al \bt}\pa_{\al}x \pa_{\bt}x
	-{1\over 2}\pa_0y^T J \pa_1y
	- {1\over 2}\pa_1 y^T J M J^T \pa_1 y			\label{jaction}
 \ee
 where $J$ is a matrix defined by
 \be
 	J = \left[I - (LML)({K_V K_V^T\over K_V^2}
				+ {K_A K_A^T\over K_A^2})\right] L.
\ee
Since $K_V$ and $K_A$ are constants and satisfy, using (\ref{bklkr}),
the relations $K_V = ML K_A$ and $K_A = ML K_V$, therefore $J$ is a
constant matrix. For the background (\ref{gb}), since
\be
	K_V = \left(\matrix {{1}\cr {1}\cr {-1}\cr {1}\cr} \right)
		\;\;\;\;\;\;
	K_A = \left(\matrix {{-1}\cr {1}\cr {1}\cr {1}\cr} \right),
\ee
we find $J = K_1K_2^T + K_2K_1^T$, where
\be
	K_1 = {1\over 2}\left(\matrix {{1}\cr {-1}\cr {1}\cr {1}\cr} \right)
			\;\;\;\;\;\;
	K_2 = {1\over 2}\left(\matrix {{1}\cr {1}\cr {1}\cr {-1}\cr} \right).
\ee
The Lagrangian (\ref{jaction}) can then be rewritten as,
\bq
	{\cL} =  -{1\over 2}\eta^{\al \bt}\pa_{\al}x \pa_{\bt}x
		-{1\over 2}(\matrix{{\pa_0 z},& {\pa_0 \tz}}) L_2
				\left(\matrix{{\pa_1 z} \cr {\pa_1 \tz}}\right)	\cr
		 -{1\over 2}\left(\matrix{{\pa_1 z},& {\pa_1 \tz}}\right)
		(L_2 M_2 L_2)
		\left(\matrix{{\pa_1 z} \cr {\pa_1 \tz} }\right)	\cr
\eq
where $z = K_1^T y$, $\tz = K_2^T y$,
\be
	L_2 = \left(\matrix{{0} & {1} \cr
					{1} & {0} \cr} \right),
		\;\;\;\;and\;\;\;\;
	M_2 = \left(\matrix{{K_1^T M K_1} & {K_1^T M K_2} \cr
					{K_2^T M K_1} & {K_2^T M K_2} \cr} \right).
\ee
For our example, i.e.  background (\ref{gb}), we find
\be
	M_2 = \left(\matrix{ {{1+b}\over {1-b}} & 0 \cr
					{0} & {{1-b}\over {1+b}} \cr}\right), \label{m2}
\ee
which is the expected form for the matrix $M$ in two dimensions
,($d=1$), for the class of background we are studying.
By setting $b=coshx$ in eqn.(\ref{gb}), which corresponds to the $SL(2, R)$
WZW model, we get
the two dimensional black hole solution from eqn.(\ref{m2}).

To summarize, in this paper we have presented the gauging procedure
for the T-duality invariant string action. We have shown the
connection of the T-duality invariant gauged actions with the
conventional gauged string actions for all  the known types of
gauging in two dimensions. We have also presented the gauge field
integration for a particular background
and shown that the resulting  action is of the
same form as written down by Schwarz and Sen. There are several
directions in which these results can be extended. First, it
should be straightforward to generalize these results to the more
general backgrounds by having more than just a single coordinate
dependence and by keeping the
off-diagonal blocks in the background metric and
antisymmetric tensor. It should also be possible to carry out
the gauge field integration  for
backgrounds other than in eqn.(\ref{gb}).  One can also
examine whether any of these results generalize for the case of
nonabelian \cite{quevedo} gaugings. We hope to report on these issues
in future.

ACKNOWLEDGEMENTS: This work was carried out mostly at ICTP, Trieste
and partly at the Institut de Physique, Neuchatel. I would like to thank
these centers for their hospitality. I would also like to thank K.
Narain and F. Quevedo for discussions.

\vfil
\eject

\vfil
\eject
\end{document}